\documentclass[11pt]{article}

\usepackage[round]{natbib}
\usepackage{wordlike}
\usepackage{setspace}

\usepackage{hyperref}
\hypersetup{colorlinks,urlcolor=blue}
\usepackage{float}
\usepackage{pdflscape}
\usepackage{textcomp}
\usepackage{times}
\usepackage{latexsym}
\usepackage{hyperref}
\usepackage{booktabs}
\usepackage{amsmath}
\usepackage{threeparttable}
\usepackage{tabularx}
\usepackage{graphicx}
\usepackage{multirow}
\usepackage{longtable}
\usepackage{xcolor}
\usepackage{threeparttable}
\usepackage{multirow}

\usepackage{amssymb}
\usepackage{upgreek}

\usepackage{url}

\usepackage{stfloats}
\usepackage{capt-of}
\definecolor{newcolor}{rgb}{.8,.349,.1}

\title{COVID-19-CT-CXR: a freely accessible and weakly labeled chest X-ray and CT image collection on COVID-19 from biomedical literature}

\author{
	\small
	Yifan Peng$^{1,3}$, Yu-Xing Tang$^{2}$, Sungwon Lee$^{2}$, Yingying Zhu$^{2,4}$, Ronald  M. Summers$^{2}$, Zhiyong Lu$^{1,*}$\\[2em]
	1. National Center for Biotechnology Information (NCBI), National Library of Medicine (NLM), National Institutes of Health (NIH), Bethesda, MD 20894\\
	2. Imaging Biomarkers and Computer-Aided Diagnosis Laboratory, Radiology and Imaging Sciences Department, National Institutes of Health (NIH) Clinical Center, Bethesda, MD 20892\\
	3. Department of Population Health Sciences, Weill Cornell Medicine, New York, NY 10065\\
	4. Department of Computer Science and Engineering, University of Texas at Arlington, Arlington, TX 76019\\[1em]
	*. Corresponding author: \url{zhiyong.lu@nih.gov}
}

\date{}

\begin{document}
	
\maketitle

\begin{abstract}
%%%
The latest threat to global health is the COVID-19 outbreak. Although there exist large datasets of chest X-rays (CXR) and computed tomography (CT) scans, few COVID-19 image collections are currently available due to patient privacy. At the same time, there is a rapid growth of COVID-19-relevant articles in the biomedical literature, including those that report findings on radiographs. Here, we present COVID-19-CT-CXR, a public database of COVID-19 CXR and CT images, which are automatically extracted from COVID-19-relevant articles from the PubMed Central Open Access (PMC-OA) Subset. We extracted figures, associated captions, and relevant figure descriptions in the article and separated compound figures into subfigures. Because a large portion of figures in COVID-19 articles are not CXR or CT, we designed a deep-learning model to distinguish them from other figure types and to classify them accordingly. The final database includes 1,327 CT and 263 CXR images (as of May 9, 2020) with their relevant text. To demonstrate the utility of COVID-19-CT-CXR, we conducted four case studies. (1) We show that COVID-19-CT-CXR, when used as additional training data, is able to contribute to improved deep-learning (DL) performance for the classification of COVID-19 and non-COVID-19 CT. (2) We collected CT images of influenza, another common infectious respiratory illness that may present similarly to COVID-19, and fine-tuned a baseline deep neural network to distinguish a diagnosis of COVID-19, influenza, or normal or other types of diseases on CT. (3) We fine-tuned an unsupervised one-class classifier from non-COVID-19 CXR and performed anomaly detection to detect COVID-19 CXR. (4) From text-mined captions and figure descriptions, we compared 15 clinical symptoms and 20 clinical findings of COVID-19 vs. those of influenza to demonstrate the disease differences in the scientific publications. Our database is unique, as the figures are retrieved along with relevant text with fine-grained descriptions, and it can be extended easily in the future. We believe that our work is complementary to existing resources and hope that it will contribute to medical image analysis of the COVID-19 pandemic. The dataset, code, and DL models are publicly available at~\url{https://github.com/ncbi-nlp/COVID-19-CT-CXR}.
%%%%
\end{abstract}

%\linenumbers

\vspace*{2em}

%% main text
\section{Introduction}

The latest threat to global health is the ongoing outbreak of the COVID-19 caused by SARS-CoV-2~\citep{fauci2020covid}. So far, pneumonia appears to be the most frequent and serious manifestation, and major complications, such as acute respiratory distress syndrome (ARDS), can present shortly after the onset of symptoms, contributing to the high mortality rate of COVID-19~\citep{chen2020epidemiological, guan2020clinical, wang2020clinical}. Chest X-rays (CXR) and chest computed tomography (CT) scans are playing a major part in the detection and monitoring of these respiratory manifestations. In some cases, CT scans have shown abnormal findings in patients prior to the development of symptoms and even before the detection of the viral RNA~\citep{shi2020radiological, xie2020chest,mei2020artificial}.

With the shortage of specialists who have been trained to accumulate experiences with COVID-19 diagnosis, there has been a concerted move toward the adoption of artificial intelligence (AI), particularly deep-learning-based methods, in COVID-19 pandemic diagnosis and prognosis, in which well-annotated data always play a critical role~\citep{shi2020review}. Although there exist large public datasets of CXR~\citep{irvin2019chexpert, johnson2019mimic, wang2017chestx} and CT~\citep{yan2018deeplesion}, there are few collections of COVID-19 images to effectively train a deep neural network~\citep{cohen2020covid, he2020sample, zhang2020clinically}.  Nevertheless, we have seen a growing number of COVID-19 relevant articles in PubMed~\citep{chen2020keep, wang2020cord}. In addition, there is a recent COVID-19 initiative to expand access via PubMed Central Open Access (PMC-OA) Subset to coronavirus-related publications and associated data (\url{https://www.ncbi.nlm.nih.gov/pmc/about/covid-19-faq/}). As a result, more articles ($>10,000$ as of May 9, 2020) relevant to the COVID-19 pandemic or prior coronavirus research were added through PMC-OA with a free-reuse license for secondary analysis. 

Non-textual components (e.g., figures and tables) provide key information in many scientific documents and are considered in many tasks, including search engine and knowledge base construction~\citep{choudhury2013figure, smith2018mouse}. As such, we have recently seen a growing interest in mining figures within scientific documents~\citep{ahmed2016mining, li2019figure, siegel2018extracting}. In the medical domain, figures also are a topical interest because they often contain graphical images, such as CXR and CT~\citep{lopez2013framework, tsutsui2017data}. Extracting CXR and CT from biomedical publications, however, is neither well studied nor well addressed. 

For the above reasons, there is an unmet need to construct the COVID-19 image dataset from PMC-OA to allow researchers to freely access the images along with a description of the text. In this paper, we thus introduce an effective framework to construct a CXR and CT database from PMC-OA and propose a public database, termed COVID-19-CT-CXR. In contrast to previous approaches that relied solely on the manual submission of medical images to the repository, in this work, figures are automatically collected by using the integration of medical imaging and natural-language processing with limited human annotation efforts. In addition, figures in this database are partnered with text that describes these cases with details, a feature not found in other such datasets. 

The framework consists of three steps. First, we extracted figures, associated captions, and relevant figure descriptions in the PMC-OA article. Such extraction is non-trivial due to the diverse layout and large volume of articles in the PMC-OA subset. Second, we separated compound figures into subfigures, as medical figures often comprise multiple image panels~\citep{li2019figure, tsutsui2017data}. Third, we classified subfigures into CXR, CT, or others because a large portion of figures in COVID-19 articles are not CXR or CT. To this end, we designed a deep-learning model to distinguish them from other figure types and to classify them accordingly.

We further demonstrate the utility of COVID-19-CT-CXR through a series of case studies. First, using this database as additional training data, we show that existing deep neural networks can receive benefits in the task of COVID-19/non-COVID-19 classification of CT images. Second, we demonstrate that the database can be used to develop a baseline model to distinguish COVID-19, influenza, and other CT, a less-studied topic. Third, we train an unsupervised one-class classifier from non-COVID-19 CXRs and performed anomaly detection to detect COVID-19 CXRs. Fourth, we extract symptoms and clinical findings from the text, using the natural language-processing methods. The symptoms and clinical findings not only confirm the results that radiologists have found but also potentially identify other findings that may have been overlooked.  

The remainder of the paper is organized as follows. Section~\ref{sec:material} presents the material and methods to build the dataset. Section~\ref{sec:results} contains the details of the statistics of the dataset, results of the image type classification, and the use cases.  Finally, Sections~\ref{sec:discussion} and~\ref{sec:conclusions} provide the discussion, conclusions, and recommendations for future work.

\section{Material and methods}
\label{sec:material}

\subsection{COVID-19 relevant articles on PMC-OA}

Articles in this study were collected from the PMC-OA Subset. PubMed Central\textsuperscript{\textregistered} (PMC) is a free, full-text archive of biomedical and life sciences journal literature (\url{https://www.ncbi.nlm.nih.gov/pmc/}). PMC-OA is a well-known portion of the PMC articles under a Creative Commons license (or custom license of the Public Health Emergency COVID-19 Initiative in PMC due to the COVID pandemic) that allows for text mining, secondary analysis, and other types of reuse (\url{https://www.ncbi.nlm.nih.gov/pmc/about/covid-19-faq/}). In this study, we collected COVID-19 relevant articles using LitCovid~\citep{chen2020keep}, a curated literature hub for tracking up-to-date scientific information about the 2019 novel coronavirus. LitCovid screens the search results of the PubMed query: \texttt{"coronavirus"[All Fields] "ncov"[All Fields] OR "cov"[All Fields] OR "2019-nCoV"[All Fields] OR "COVID-19"[All Fields] OR "SARS-CoV-2"[All Fields]}. Relevant articles are identified and curated with assistance from an automated machine-learning and text-classification algorithm. As of May 9, 2020, there were 5,381 PMC-OA articles in the collection (Table~\ref{tab:overview}). The topics of articles ranged from diagnosis to treatment to case reports. 
\begin{table}[!htbp]
	\caption{An overview of the COVID-19 relevant articles as of May 9, 2020.}
	\label{tab:overview}
	\centering
	\begin{tabular}{lr}
		\toprule
		Characteristics & $n$\\
		\midrule
		COVID-19 relevant articles in PMC-OA & 5,381\\
		\hspace{1em}Prevention & 2,089\\
		\hspace{1em}Mechanism & 577\\
		\hspace{1em}Diagnosis & 546\\
		\hspace{1em}Case Report & 355\\
		\hspace{1em}Transmission & 354\\
		\hspace{1em}General & 238\\
		\hspace{1em}Epidemic Forecasting & 64\\
		\hspace{1em}Others & 1,158\\
		Journals & 1,145\\
		Figures & 4,407\\
		\bottomrule
	\end{tabular}
\end{table}

\subsection{Overview of the COVID-19-CT-CXR construction}

\begin{figure*}[!htbp]
	\centering
	\includegraphics[width=.8\textwidth, clip, trim=0 160 220 0]{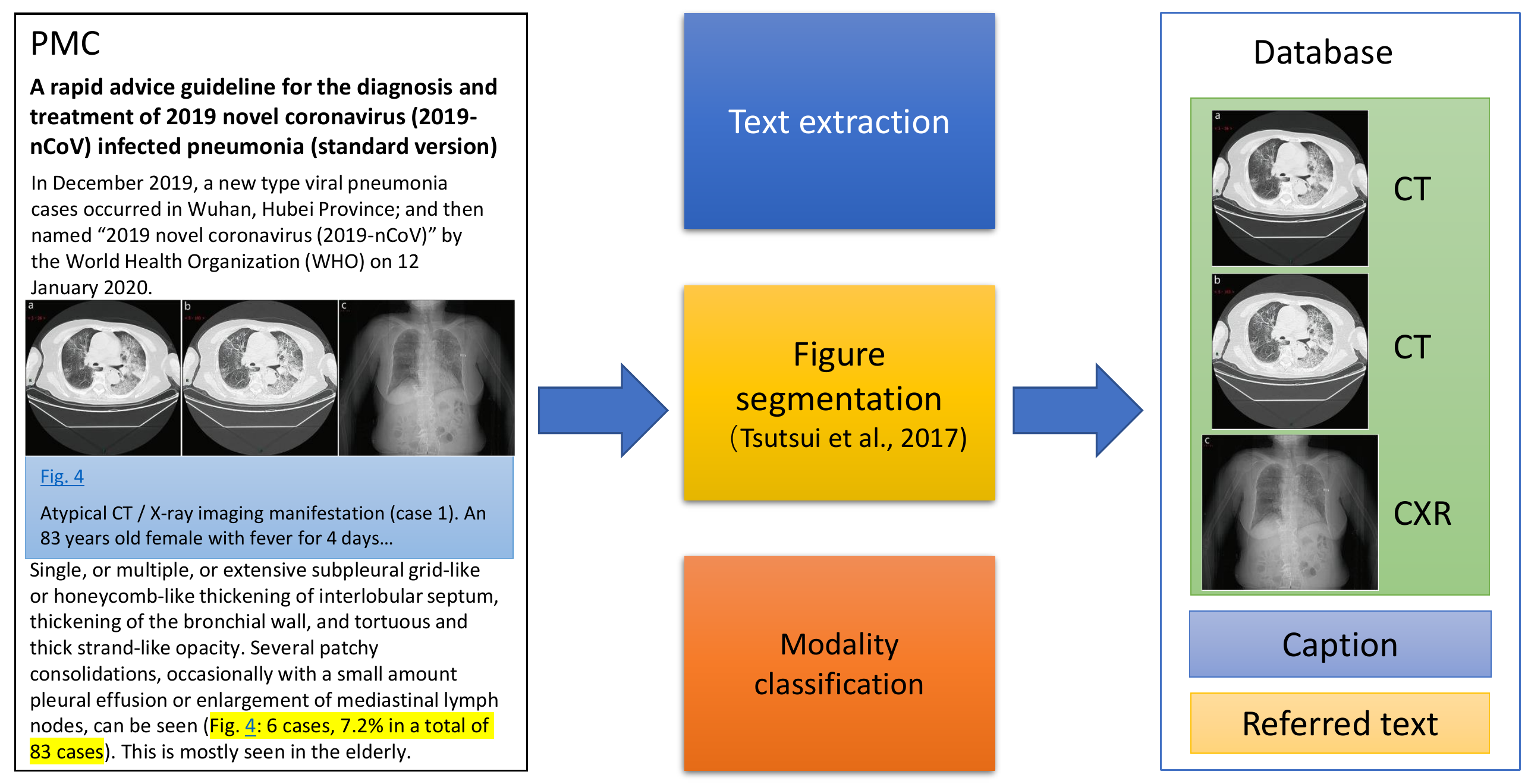}
	\caption{The overview of the pipeline to collect the images with text.}
	\label{fig:overview}
\end{figure*}

Figure~\ref{fig:overview} shows the overview pipeline of the development. For a given PMC-OA article, we first extract figures, associated captions, and relevant figure descriptions in the PMC-OA article. Then, if figures are compound, we separate them into subfigures. We further classify the individual figures into CT, CXR, or other types of scientific images, using a deep-learning model. The final database includes figures with their types and relevant descriptions in the manuscript. 

\subsection{Text extraction}

\begin{figure}[!htbp]
	\centering
	\includegraphics[width=\columnwidth]{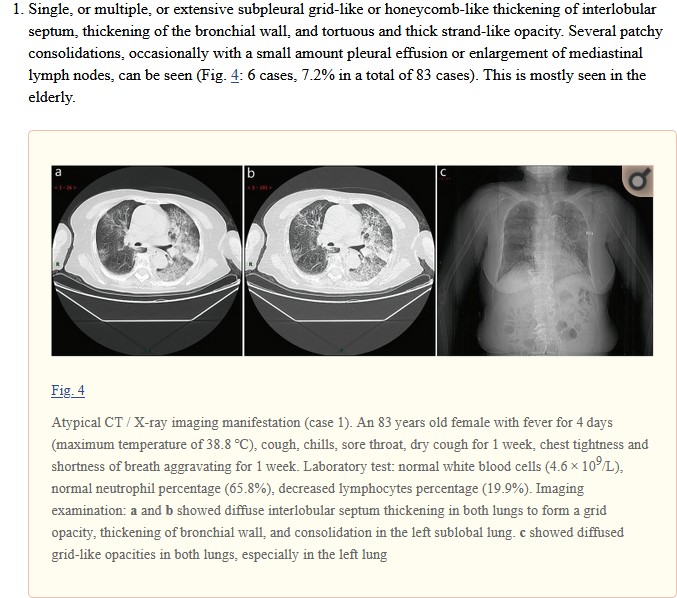}
	\caption{Examples of CT and CXR that are positive for COVID-19. The figures are from the article, ``A rapid advice guideline for the diagnosis and treatment of 2019 novel coronavirus (2019-nCoV) infected pneumonia (standard version)''~\citep{jin2020rapid}.}
	\label{fig:example}
\end{figure}

In this step, we identify figure captions and relevant text with the referenced figures. To facilitate the automated processing of full-text articles in PMC-OA, \cite{comeau2019pmc} convert PMC articles to BioC format, a data structure in XML for text sharing and processing. Each article in BioC format is encoded in UTF-8, and Unicode characters are converted to strings of ASCII characters. The article also includes section types, figures, tables, and references~\citep{kafkas2015section}. In this study, we downloaded the PMC-OA articles through the RESTful web service (\url{https://www.ncbi.nlm.nih.gov/research/bionlp/APIs/BioC-PubMed/}). We parsed these articles to locate figures with their figure numbers and their captions. We then used the figure number and regular expressions to find where the figure is cross-referenced in the document. Figure~\ref{fig:example} shows an example of a typical biomedical image in the article, ``A rapid advice guideline for the diagnosis and treatment of 2019 novel coronavirus (2019-nCoV) infected pneumonia (standard version)''~\citep{jin2020rapid}. The examples contain CXR, CT, a figure caption, and text that describes the case with rich information, such as fever, symptoms, and clinical findings.

\subsection{Subfigure separation}

Most of the figures in the PMC-OA articles are compound figures. A key challenge here is that one figure may have individual subfigures of the same category (e.g., four CT images) or several categories (e.g., one CXR and one CT image placed side by side). For example, Figure~\ref{fig:example} contains a compound figure with three subfigures~\citep{jin2020rapid}. Figures~\ref{fig:example}a and Figure~\ref{fig:example}b are CT images, and Figure~\ref{fig:example}c is a CXR. Notably, it is a requirement to decompose compound figures into subfigures before modality classification. In this study, we used a convolutional neural network developed by~\citet{tsutsui2017data} to separate compound figures. The model was pretrained on the ImageCLEF Medical dataset with an accuracy of 85.9\%~\citep{de2016overview}. 

We applied the model on the figures obtained in previous steps and filtered the subfigures with a size smaller than 224 x 224 pixels. We consider that subfigures with fewer pixels might be deformed, and most state-of-the-art neural networks in image analysis, such as Inception-v3~\citep{szegedy2016rethinking} and DenseNet~\citep{iandola2014densenet}, require an input size of 224 or larger.

\subsection{Image modality classification}

A large portion of figures in the PMC-OA articles are not CXR or CT images. To distinguish them from other types of scientific figures, we designed a scientific figure classifier that was fine-tuned on a newly created dataset (\url{https://github.com/ncbi-nlp/COVID-19-CT-CXR}). Table~\ref{tab:summary} shows the breakdown of the figures by their category in the training and test set. This dataset consists of 2,700 figures in three categories: CXR, CT, and Other scientific figure types. A total of 500 CXRs are randomly picked from the NIH Chest X-ray~\citep{wang2017chestx}, and 500 CT images are randomly picked from DeepLesion~\citep{yan2018deeplesion}. Other scientific figures are randomly picked from DocFigure~\citep{jobin2019docfigure}. The original DocFigure annotated figures of 28 categories, such as Heat map, Bar plots, and Histogram. Here, we combined these categories into one for simplicity of training the classifier. In addition, we curated 1,200 figures from PMC-OA, using the annotation tool developed by  \citet{tang2020automated}. 

\begin{table}[!htbp]
	\caption{Summary of the dataset for image modality classification.}
	\label{tab:summary}
	\centering
	\begin{tabular}{lrr}
		\toprule
		Modality & Training & Test\\
		\midrule
		CXR\\
		\hspace{1em}NIH Chest X-ray~\cite{wang2017chestx} & 399 & 101\\
		\hspace{1em}PMC-OA & 38 & 7\\
		CT		\\
		\hspace{1em}DeepLesion~\cite{yan2018deeplesion} & 415 & 85\\
		\hspace{1em}PMC-OA & 225 & 21\\
		Other scientific document figures		\\
		\hspace{1em}DocFigure~\cite{jobin2019docfigure} & 386 & 114\\
		\hspace{1em}PMC-OA & 737 & 172\\
		\midrule
		Total & 2,200 & 500\\
		\bottomrule
	\end{tabular}
\end{table}

Our framework uses DenseNet121 to classify image types~\citep{huang2016densely}. The weights (or parameters) were pretrained on ImageNet~\citep{russakovsky2015imagenet}. We replaced the last classification layer with a fully connected layer with a softmax operation that outputs the approximate probability that an input image is a CXR, CT, or other scientific figure type. All images were resized to 224 x 224 pixels. The hyperparameters include a learning rate of 0.0001, a batch size of 16, and 50 training epochs. All experiments were conducted on a server with an NVIDIA V100 128G GPU from the NIH HPC Biowulf cluster (\url{http://hpc.nih.gov}). We implemented the framework using the Keras deep-learning library with TensorFlow backend (\url{https://www.tensorflow.org/guide/keras}).

\subsection{Qualification and statistical analysis}

The performance metrics include the area under the receiver operating characteristic curve (AUC), sensitivity, specificity (recall), precision (positive predictive value), and F1 score. For the classification problem, we chose the label with the highest probability when required in computing the metrics. Each of the models was fine-tuned and tested five times, using the same parameters, training, and testing images each time. The validation set was randomly selected from 10\% of the training set. Fisher's exact test was used to determine whether there are nonrandom associations between COVID-19 and influenza's symptoms and clinical findings~\citep{fisher1922interpretation}. We conduct above statistical analysis using numpy, scipy, matplotlib, and scikit-learn built on Python.

\section{Results}
\label{sec:results}

\subsection{COVID-19-CT-CXR characteristics}

\begin{figure}[!htbp]
	\centering
	\includegraphics[width=.6\columnwidth, trim=1em 1em 4em 2em,clip]{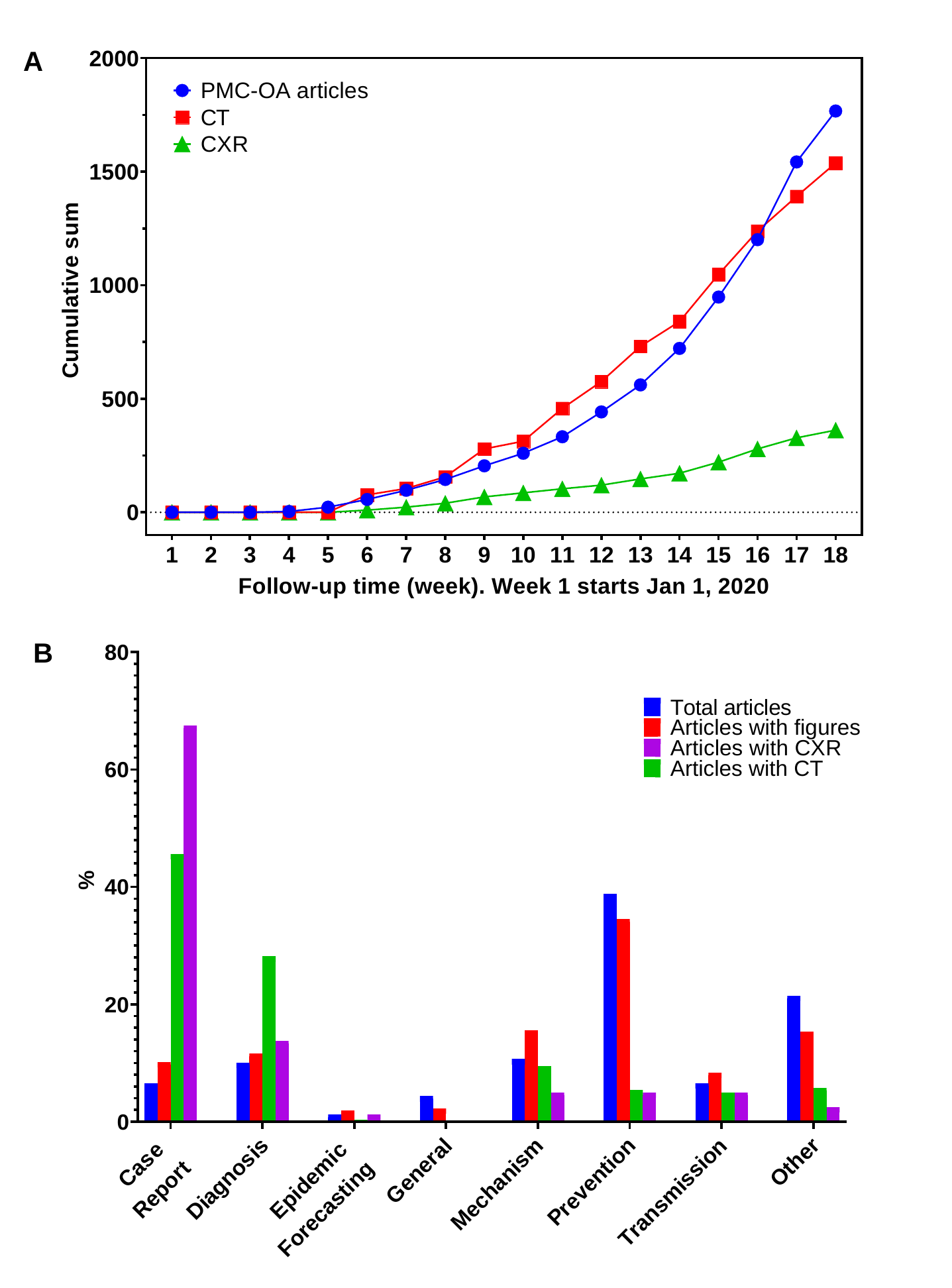}
	\caption{Characteristics of the COVID-19-CT-CXR. (A) The rapid growth of the number of COVID-19-relevant articles, CT, and CXR in PMC-OA from January 1, 2020 (Week 1). (B) The distribution of categories in COVID-19-relevant PMC-OA articles and articles with figures, CT, and CXR.}
	\label{fig:characteristics}
\end{figure}

Table~\ref{tab:summary covid-19-ct-cxr} shows the breakdown of the figures by modality. We obtained 1,327 CT images and 263 CXR text-mined labeled as positive for COVID-19 from 1,831 PMC-OA articles. These images have different sizes. The minimum, maximum, and average heights are 224, 2,703, and 387.5 pixels, respectively. The minimum, maximum, and average widths are 224, 1,961, and 472.4, respectively. For each article, we also include major elements, such as DOI, title, journal, and publication date for reference. Figure~\ref{fig:characteristics}A shows the cumulative numbers of articles and figures on a weekly basis. We analyzed the proportional distribution of categories in COVID-19 relevant PMC-OA articles, and articles with figures, CT, and CXR. Figure~\ref{fig:characteristics}B shows that the “Case Report” category contains higher proportional articles with CXR/CT.

\begin{table}[!htbp]
	\caption{Summary of the COVID-19-CT-CXR dataset}
	\label{tab:summary covid-19-ct-cxr}
	\centering
	\begin{tabularx}{.47\textwidth}{Xr}
		\toprule
		Characteristics & $n$\\
		\midrule
		PMC-OA articles with figures & 1,831\\
		Subfigures & 10,650\\
		\hspace{1em}CXR & 263\\
		\hspace{1em}CT & 1,327\\
		\hspace{1em}Others & 9,060\\
		\bottomrule
	\end{tabularx}
\end{table}

\subsection{Image modality classification}

Table~\ref{tab:image type} shows the performance of the model to classify image modality. The macro average \textit{F}-score is 0.996. The \textit{F}-score was 0.993 ± 0.004 for CT, 1.000 ± 0.000 for CXR, and 0.998 ± 0.001 for other scientific figure types. 

\begin{table}[!hbtp]
	\caption{The performance of image type classification. The test set is the combination of NIH Chest X-ray, DeepLesion, DocFigure, and PMC-OA.}
	\label{tab:image type}
	\centering
	\begin{tabular}{lcccc}
		\toprule
		Metrics & CT & CXR & Other scientific figures & \textit{Macro Avg}\\
		\midrule
		Precision          & 0.989 ± 0.004 & 1.000 ± 0.000 & 0.999 ± 0.001 & 0.996 ± 0.002\\
		Recall/Sensitivity & 0.998 ± 0.004 & 1.000 ± 0.000 & 0.996 ± 0.001 & 0.998 ± 0.002\\
		Specificity        & 0.997 ± 0.001 & 1.000 ± 0.000 & 0.999 ± 0.002 & 0.999 ± 0.001\\
		\textit{F}-score            & 0.993 ± 0.004 & 1.000 ± 0.000 & 0.998 ± 0.001 & 0.997 ± 0.002\\
		\bottomrule
	\end{tabular}
\end{table}

\subsection{Use cases}

To demonstrate the utility of COVID-19-CT-CXR, we conducted four case studies. (1) We combined COVID-19-CT-CXR with previously curated data at \url{https://github.com/UCSD-AI4H/COVID-CT}~\cite{zhao2020covid} and fine-tuned a deep neural network to perform the classification of COVID-19 and non-COVID-19 CT. (2) We collected CT of influenza, using a similar method, and fine-tuned a deep neural network to distinguish among the diagnoses of COVID-19, influenza, and normal or other types of diseases on CT. (3) We fine-tuned an unsupervised one-class learning model, using only non-COVID-19 CXR to perform anomaly detection, to detect COVID-19 CXR. (4) We extracted 15 clinical symptoms and 26 clinical findings from the captions and relevant descriptions. We then compared their frequencies to those described in articles on influenza, another common infectious respiratory illness that may present similarly to COVID-19.

\subsubsection{Classification of COVID-19 and non-COVID-19 on CT}

In the context of the COVID-19 pandemic, it is important to separate patients likely to be infected with COVID-19 from other non-COVID-19 patients. As it is time-consuming for specialists to both accumulate experiences and read a large volume of CT scans to diagnose COVID-19, many studies use machine learning to separate COVID-19 patients from non-COVID-19 patients~\citep{chen2020deep, he2020sample, jin2020development, wang2020deep, zheng2020deep}. In this work, we hypothesize that our creation of additional training data from existing articles can improve the performance of the system and reduce the effort of manual image annotation. To test this hypothesis, we compared the performance of deep neural networks fine-tuned on the existing benchmark~\citep{zhao2020covid} and COVID-19-CT-CXR (Table~\ref{sup tab:summary of 2-class}). For a fair comparison, we added additional training examples only in the training set and used the same test set as described in~\citet{he2020sample}.

\begin{table}[!htbp]
	\caption{Summary of the dataset for classification of COVID-19 and non-COVID-19 CT.}
	\label{sup tab:summary of 2-class}
	\centering
	\begin{tabular}{llrr}
		\toprule
		Dataset &  & COVID-19 & Non-COVID-19\\
		\midrule
		Training & \cite{zhao2020covid} & 251 & 292\\
		& COVID-19-CT & 542 & 67\\
		Test & \cite{zhao2020covid} & 98 & 105\\
		\bottomrule
	\end{tabular}
\end{table}

\begin{figure}[!htbp]
	\centering
	\includegraphics[width=.8\columnwidth,clip,trim=1em 1em 1em 1em]{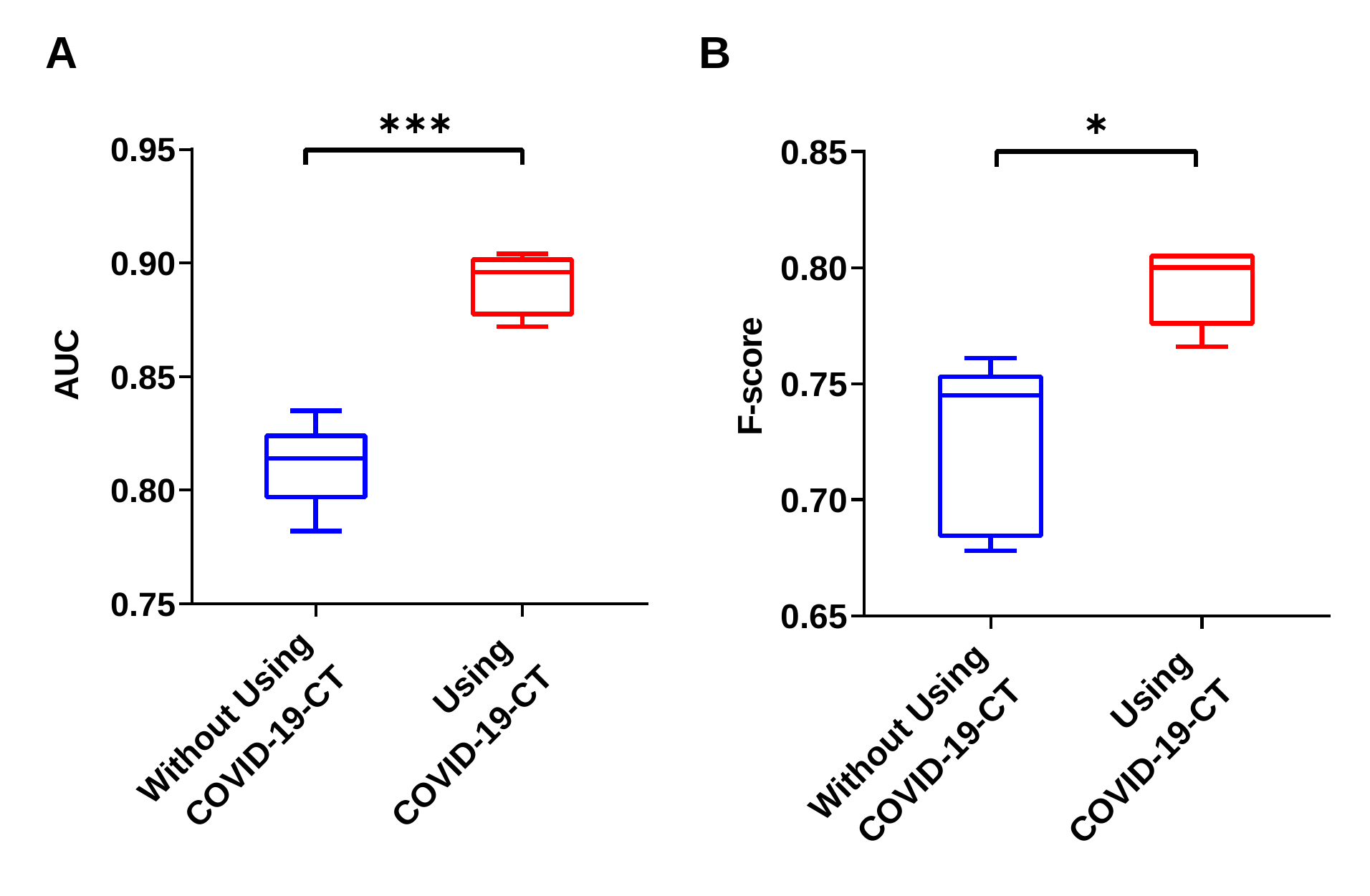}
	\caption{Comparison of AUC and \textit{F}-score by models fine-tuned with and without using additional COVID-19 CT extracted from PMC-OA. *: $P \leq 0.05$; ***: $P \leq 0.001$ (t-test).}
	\label{fig:auc}
\end{figure}

In this experiment, DenseNet121 was pre-trained on ImageNet, fine-tuned, and evaluated on the training and test sets. We then replaced the last classification layer with a single neuron with sigmoid that outputs the approximate probability that an input image is COVID-19 or non-COVID-19. Other experimental settings are the same as that of fine-tuning the image modality classifier. Figure~\ref{fig:auc} shows that the model significantly outperforms the baseline when PMC-OA CT figures were added for fine-tuning. Specifically, we achieved the highest performance of 0.891 ± 0.012 in AUC, 0.780 ± 0.074 in recall, 0.816 ± 0.053 in precision, and 0.792 ± 0.015 in \textit{F}-score (Table~\ref{sub tab:performance of 2-class}).

\begin{table}[!htbp]
	\caption{Performance metrics for classification of COVID-19 and non-COVID-19 CT.}
	\label{sub tab:performance of 2-class}
	\centering
	\begin{tabular}{lcc}
		\toprule
		Metrics & Without using COVID-19-CT & Using COVID-19-CT\\
		\midrule
		AUC                & 0.811 ± 0.017 & 0.891 ± 0.012\\
		Precision          & 0.742 ± 0.029 & 0.816 ± 0.053\\
		Recall/Sensitivity & 0.714 ± 0.083 & 0.780 ± 0.074\\
		Specificity        & 0.764 ± 0.059 & 0.827 ± 0.073\\
		\textit{F}-score            & 0.724 ± 0.034 & 0.792 ± 0.015\\
		\bottomrule
	\end{tabular}
\end{table}

\subsubsection{Classification of COVID-19, influenza, and other types of disease on CT}

As the COVID-19 outbreak continues to evolve, there is an increasing number of studies that compare COVID-19 with other viral pneumonias, such as influenza~\citep{luo2020using}. Distinguishing patients infected by COVID-19 and influenza is important for public health measures because the current treatment guidelines are different~\citep{kimberlin2018reda}. This task is non-trivial because both viruses have a similar radiological presentation. To assist clinicians at triage, several studies have proposed to use deep learning to distinguish COVID-19 from influenza and no-infection with 3D CT scans~\citep{xu2020deep}. In this paper, we aim to establish a baseline model to distinguish COVID-19 from influenza on single CT figures. To collect CT figures with influenza, we searched the PMC using the query ``\texttt{(Influenza[Title] OR (flu[Title] AND pneumonia[Title]) AND open access[Filter]}'' and extracted the most recent 10,000 PMC-OA articles. We used the same method to extract CT and its caption and relevant text from the articles (called Influenza-CT). Taken together, we construct a dataset with 983 CT for training and 242 CT for testing (Table~\ref{sub tab:summary of 3-class}).

\begin{table}[!hbtp]
	\caption{Summary of the dataset for classification of COVID-19, influenza, and others in CT.}
	\label{sub tab:summary of 3-class}
	\centering
	\begin{tabular}{lccc}
		\toprule
		Dataset & COVID-19 & Influenza & Normal or other diseases\\
		\midrule
		Training & 488 & 177 & 318\\
		Test & 118 & 45 & 79\\
		\bottomrule
	\end{tabular}
\end{table}

\begin{figure}[!hbtp]
	\centering
	\includegraphics[width=.6\columnwidth,clip,trim=1em 0 4em 3em]{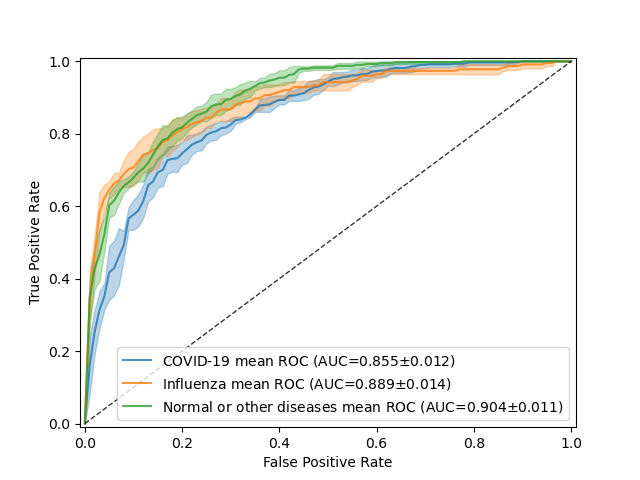}
	\caption{Receiver operating characteristic (ROC) curves of the classification of COVID-19, influenza, and normal or other types of diseases in CT. The model was fine-tuned and tested 5 times, using the same training and testing images each time. The mean ROC curve is shown together with its standard deviation (shaded area).}
	\label{fig:roc}
\end{figure}

To obtain the baseline model, we use the same model and experimental settings as described in the ``Image modality classification'' section. Figure~\ref{fig:roc} shows the performance of the deep-learning model by its receiver operating characteristic (ROC) curves. The AUC was 0.855 ± 0.012 for COVID-19 detection and 0.889 ± 0.014 for influenza detection. Table \ref{sub tab:performance of 3-class} shows more detail for the results. We achieved the highest precision (0.845 ± 0.026) for COVID-19 detection and high recall (0.711 ± 0.053) for influenza detection.

\begin{table}[!hbtp]
	\caption{Performance metrics for classification of COVID-19, influenza, and normal or other types of diseases in CT.}
	\label{sub tab:performance of 3-class}
	\begin{center}
		\begin{tabular}{lcccc}
			\toprule
			Metrics & COVID-19 & Influenza & Normal or other diseases & \textit{Macro Avg}\\
			\midrule
			AUC                & 0.855 ± 0.012 & 0.889 ± 0.014 & 0.904 ± 0.011 & 0.879 ± 0.010\\
			Precision          & 0.845 ± 0.026 & 0.609 ± 0.033 & 0.642 ± 0.021 & 0.699 ± 0.019\\
			Recall/Sensitivity & 0.597 ± 0.030 & 0.711 ± 0.053 & 0.861 ± 0.033 & 0.723 ± 0.022\\
			Specificity        & 0.895 ± 0.024 & 0.895 ± 0.013 & 0.767 ± 0.025 & 0.852 ± 0.009\\
			\textit{F}-score   & 0.699 ± 0.018 & 0.655 ± 0.034 & 0.735 ± 0.015 & 0.696 ± 0.018\\
			\bottomrule
		\end{tabular}
	\end{center}
\end{table}

\subsubsection{Anomaly detection of COVID-19 in CXR using one-class learning}

\begin{figure}[!hbtp]
	\centering
	\includegraphics[width=.6\columnwidth,clip,trim=1em 0 4em 3em]{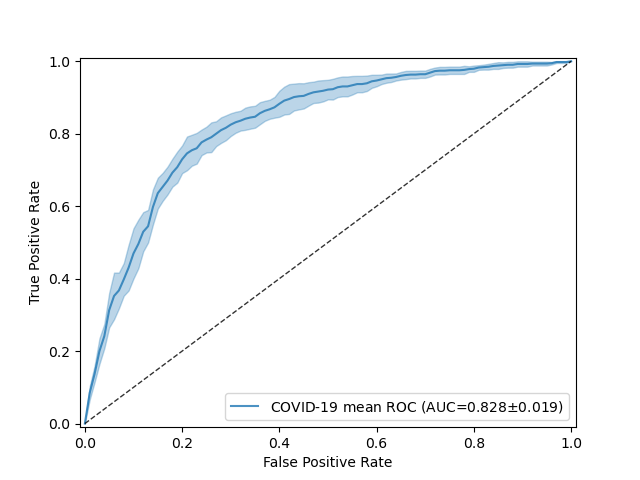}
	\caption{Receiver operating characteristic (ROC) curves of the classification of COVID-19 anomaly detection in CXR. The model was fine-tuned and tested 5 times, using the same training and testing images each time. The mean ROC curve is shown together with its standard deviation (shaded area).}
	\label{fig:one class}
\end{figure}

As they lack annotated COVID-19 CXR for training powerful deep-learning classifiers, unsupervised and semi-supervised approaches are highly desired for automated COVID-19 diagnosis. The presence of COVID-19 can be considered a novel anomaly in CXR for the NIH Chest X-ray dataset, in which no COVID-19 cases are available. In this experiment, we performed anomaly detection~\citep{chandola2009anomaly, zhang2020covid} to detect COVID-19 CXR. We trained a one-class classifier, using only non-COVID-19 CXR, and used this classifier to distinguish COVID-19 CXR from non-COVID-19 CXR. The non-COVID-19 images were a subset extracted from the NIH Chest X-ray dataset by combining 14 abnormalities and a no-finding category. The detailed numbers of training and testing CXR are shown in Table~\ref{sub tab:summary of 1-class}. We adopted the generative adversarial one-class learning approach from~\citet{tang2019abnormal}. Figure~\ref{fig:one class} shows the performance of the unsupervised one-class learning by its ROC curves. Table~\ref{sub tab:performance of 1-class} shows more detail for the results. Our model achieved 0.828 ± 0.019 in AUC, 0.767 ± 0.020 in precision, 0.772 ± 0.017 in recall, and 0.769 ± 0.018 in \textit{F}-score for COVID-19 anomaly detection.

\begin{table}[!hbtp]
	\caption{Summary of dataset used for anomaly detection of COVID-19 in CXR in unsupervised one-class classification.}
	\label{sub tab:summary of 1-class}
	\centering
	\begin{tabular}{lcc}
		\toprule
		Dataset & COVID-19 & Non-COVID-19\\
		\midrule
		Training & 0 & 37,829\\
		Test & 184 & 184\\
		\bottomrule
	\end{tabular}
\end{table}

\begin{table}[!hbtp]
	\caption{Anomaly detection performance of COVID-19 vs. non-COVID-19 using unsupervised one-class learning.}
	\label{sub tab:performance of 1-class}
	\centering
	\begin{tabular}{lc}
		\toprule
		Metrics & COVID-19 vs Non-COVID-19\\
		\midrule
		AUC                & 0.828 ± 0.019\\
		Precision          & 0.767 ± 0.020\\
		Recall/Sensitivity & 0.772 ± 0.017\\
		Specificity        & 0.765 ± 0.023\\
		\textit{F}-score   & 0.769 ± 0.018\\
		\bottomrule
	\end{tabular}
	\vspace{-1em}
\end{table}

\subsubsection{Extraction of clinical symptoms and findings using text-mining}

In this case, we extracted clinical symptoms or signs from the figure captions and relevant text that describes the case. A total of 15 symptoms or signs were collected from~\citet{guan2020clinical} and the CDC website (\url{https://www.cdc.gov/coronavirus/2019-ncov/symptoms-testing/symptoms.html}), including chest pain, constipation, cough, diarrhea, dizziness, dyspnea, fatigue, fever, headache, myalgia, proteinuria, runny nose, sputum production, throat pain, and vomiting.

\begin{figure}[!htbp]
	\centering
	\includegraphics[width=.7\columnwidth,clip,trim=1em 1em 1em 1em]{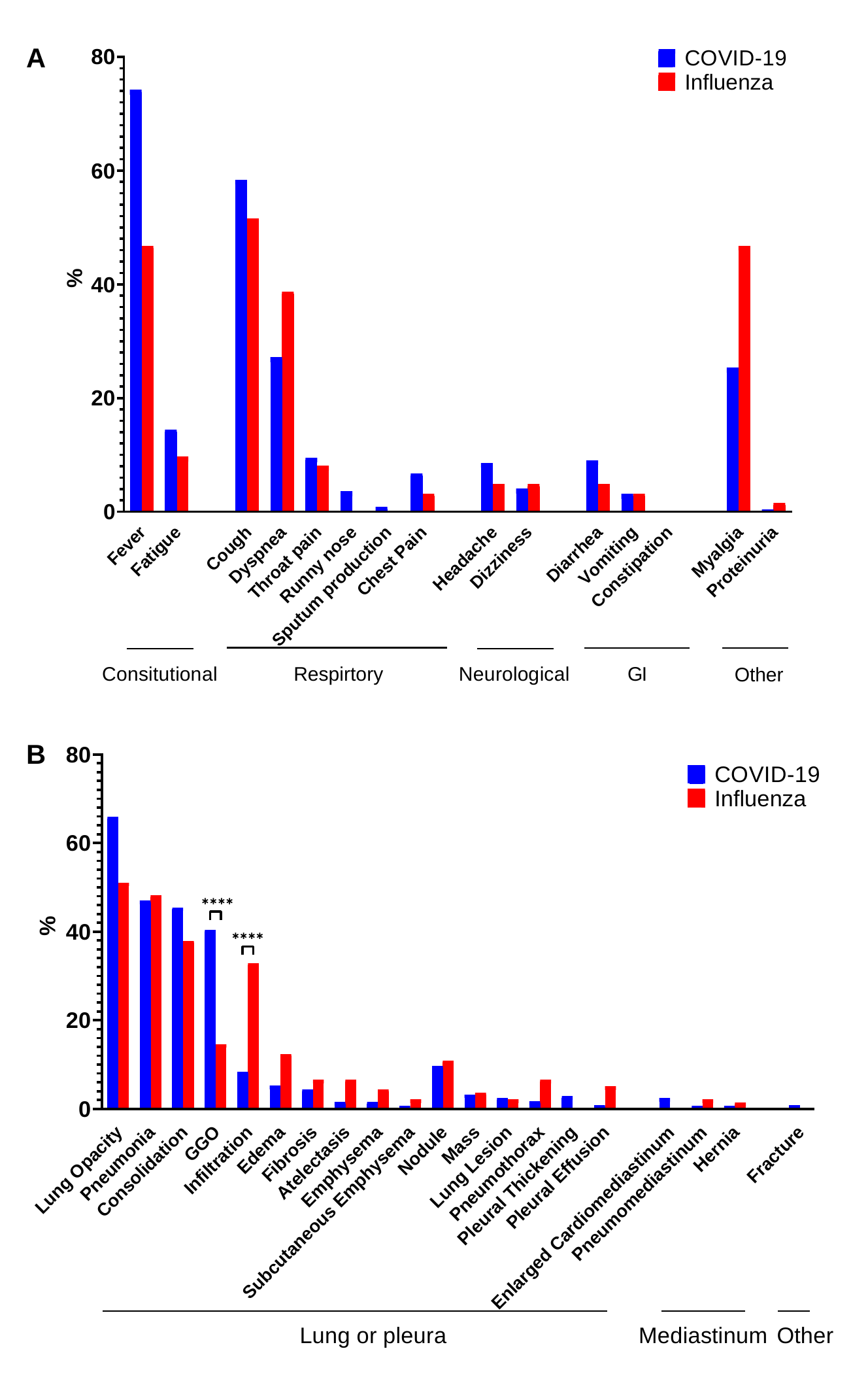}
	\caption{The frequencies of (A) 15 symptoms and (B) 20 clinical findings text mined from the figure captions and relevant text from the collection of COVID-19- and influenza-relevant articles. ****: $p \leq 0.0001$ (Fisher exact test).}
	\label{fig:findings}
	\vspace{-1em}
\end{figure}

Extracting these symptoms from text is a challenging task because their mentions in the text can be positive or negative. For example, ``fever'' is negative in the sentence, ``She experienced headache and pharyngalgia but no fever on 29 January.'' To discriminate between positive and negative mentions, we applied our previously developed tool, NegBio, on the figure caption and referred text~\citep{peng2018negbio}. In short, NegBio utilizes patterns in universal dependencies to identify the scope of triggers that are indicative of negation; thus, it is highly accurate for detecting negative symptom mentions. Figure~\ref{fig:findings}A shows the proportion of symptoms for COVID-19 and influenza. The most common symptoms are fever, cough, dyspnea, and myalgia.

We then extracted the radiographic findings from the figure caption and text. The findings (and their synonyms) are based on 20 common thoracic disease types, which are expanded from NIH Chest X-ray 14 labels~\citep{wang2017chestx}. Figure~\ref{fig:findings}B shows the 20 findings in both COVID-19 and influenza datasets. Both illnesses can result in lung opacity, pneumonia, and consolidation. COVID-19 more likely results in ground-glass opacification (GGO), while influenza more likely results in infiltration than does COVID-19 (Fisher's exact test, $p<0.0001$).

\section{Discussion}
\label{sec:discussion}

In this abrupt outbreak of SARS-CoV-2, the demand for chest radiographs and CT scans is growing rapidly, but there is a shortage of experienced specialists, radiologists, and researchers. Further, we are still new to this virus and have yet to discover the full radiologic features and prognosis of this disease. The tremendous increase in the number of patients has led to a substantial increase of COVID-19-related PMC-OA articles over the past few months (Figur~\ref{fig:characteristics}A), especially in the case report and diagnosis-relevant articles (Figure~\ref{fig:characteristics}B). These articles contain rich chest radiographs and CT images that are helpful for scientists and clinicians in describing COVID-19 cases. Thus, it is important to analyze these images and text to construct a large-scale database. By using the quickly increasing dataset, AI methods can help to find significant features of COVID-19 and speed up the clinical workload. Among others, deep learning is undoubtedly a powerful approach in dealing with a pandemic outbreak of COVID-19. 

Although deep learning has shown promise in diagnosing/screening COVID-19, using CT, it remains difficult to collect large-scale labeled imaging data, especially in the public domain. In this work, we present a set of repeatable techniques to rapidly build a CT and CXR dataset of COVID-19 from PMC-OA COVID-19-relevant articles. The strength of the study lies in its multidisciplinary integration of medical imagining and natural-language processing. It provides a new way to annotate large-scale medical images required by deep-learning models.  

An additional strength includes a highly accurate model for image type classification. As a large portion of figures in the PMC-OA articles are not CXR or CT images, we provided a model to classify these two types from other scientific figure types. Our model achieved both high precision and high recall (Table~\ref{tab:image type}). 

To assess the hypothesis that deep neural network fine-tuning on this additional dataset enables us to diagnose COVID-19 with almost no hand-labeled data, we conducted several experiments. First, we showed that this additional data enable significant performance gains to classify COVID-19 versus non-COVID-19 lung infection on CT (Figure~\ref{fig:auc} and Table~\ref{sub tab:performance of 2-class}). For our own system, we show that our baseline performance compares favorably to the results in~\citep{he2020sample}. Then, we added more automatically labeled training data and achieved the highest performance of 0.891 ± 0.012 in AUC. The comparison shows that, with additional data, both precision and recall substantially improve (7.4\% and 6.6\%, respectively). This observation indicates that additional COVID-19 CT helps to not only find more but also to restrict the positive predictions to those with the highest certainty in the model. 

In a more challenging scenario, we built a baseline system to distinguish COVID-19, influenza, and no-infection CT, which is a more clinically interesting but also more challenging task. We observed that we could achieve high AUCs for both COVID-19 and influenza detection. The recall of COVID-19 detection and the precision of influenza, however, are low (0.597 ± 0.030 and 0.609 ± 0.033, respectively). Although several studies have tackled this problem~\citep{xu2020deep}, to the best of our knowledge, there is no publicly available benchmarking. The differentiation between COVID-19 and influenza on CXR/CT without associated context is challenging. In the experiment on classification of COVID-19, influenza, and other types of disease on CT, we found that although many of the CT findings had overlapping findings, “mixed GGO (Ground glass opacity)” were mostly found in the COVID-19 dataset and “pleural thickening” and “linear opacities” were mostly found in the influenza dataset. It is also worthy to note that the images from PMC-OA may not represent the typical pool of influenza pneumonia real-world images, since researchers may report extreme cases instead of typical cases. While our work only scratches the surface of the classification of COVID-19, influenza, and normal or other types of diseases, we hope that it sheds light on the development of generalizable deep-learning models that can assist frontline radiologists.

In addition, we presented a one-class learning model for anomaly detection of COVID-19 in CXR by learning only from non-COVID-19 radiographs. Compared to the CT-based method, the one-class model achieves comparable performance, showing great potential in discriminating COVID-19 from CXR. The performance of our model, however, is worse than  that of~\cite{zhang2020covid}, suggesting that this weakly labeled dataset should be used as additional training data obtained without additional annotation cost from existing entries in curated databases.

The unique characteristic of our database is that figures are retrieved along with relevant text that describes these cases in detail. Thus, text mining can be applied to extract additional information that confirms the existing results and potentially identifies other findings that may have been overlooked. As proof of this concept, we extracted clinical symptoms and findings from the text. We found that the most common symptoms of COVID-19 were fever and cough (Figure~\ref{fig:findings}A), which are consistent with the clinical characteristics in~\cite{zhang2020clinically}. Other common symptoms include dyspnea (shortness of breath), fatigue, and throat pain. These symptoms are consistent with those reported by the CDC. When comparing the frequencies of these 20 clinical findings to those described in articles on influenza, Figure 7 shows that both conditions cause lung opacity, pneumonia, and consolidation. Further, GGO appears more frequently for COVID-19, whereas ``infiltration'' appears more frequently for influenza. This is because radiologists use the term \textit{GGO} to describe most COVID-19 findings. In addition, the influenza articles are older than are the COVID articles, and, according to Fleischner Society recommendations, the use of the term \textit{infiltrate} remains controversial, and it is recommended that it no longer be used in reports~\citep{bueno2018updated}.

In terms of limitations, first, the subfigure segmentation model needs to be improved. In this study, we applied a deep-learning model that was pretrained on an ImageCLEF Medical dataset to this task~\citep{tsutsui2017data}. Although this model is robust to variations in background color and spaces between subfigures, it sometimes fails to recognize similar subfigures that are aligned very closely. Unfortunately, these cases appear more frequently in our study than in others (e.g., several CT images are placed in a grid). Other errors occur when the model incorrectly treated the spine as spaces in the anteroposterior (AP) chest X-ray and split the large figure into two subfigures. In the future, the figure synthesis approach should be applied to augment the training datasets. Another limitation is that this work extracted only the passage that contains the referred figure. Sometimes, the case is not described in this passage. In the future, we plan to text mine the associated case description in the full text. Finally, while a figure is typically copyrighted with the original article and using previously published figures is not a common practice in scholarly publications, it is possible that one image is reused in different papers or reused in one paper for different purposes.  In the future, we plan to develop a model to remove duplicated images in the collection.

\section{Conclusions}
\label{sec:conclusions}

We have developed a framework for rapidly constructing a CXR/CT database from PMC full-text articles. Our database is unique, as figures are retrieved along with relevant text that describes these cases in detail, and it can be extended easily in the future. Hence, the work is complementary to existing resources. Applications of this database show that our creation of additional training data from existing articles improves the system performance on COVID-19 vs. non-COVID-19 classification in CT and CXR. We hope that the public dataset can facilitate deep-learning model development, educate medical students and residents, help to evaluate findings reported by radiologists, and provide additional insights for COVID-19 diagnosis. With an ongoing commitment to data sharing, we anticipate increasingly adding CXR and CT images to be made available as well in the coming months. The code that extracts the text from PMC, segments subfigures, and classifies image modality is openly available at \url{https://github.com/ncbi-nlp/COVID-19-CT-CXR}.

% if have a single appendix:
%\appendix[Proof of the Zonklar Equations]
% or
%\appendix  % for no appendix heading
% do not use \section anymore after \appendix, only \section*
% is possibly needed

% use appendices with more than one appendix
% then use \section to start each appendix
% you must declare a \section before using any
% \subsection or using \label (\appendices by itself
% starts a section numbered zero.)
%
%
%
%\appendices
%\beginsupplement
%\noindent
%% you can choose not to have a title for an appendix
%% if you want by leaving the argument blank
%\section{}
%
%%
%
%
%\vspace{1em}
%
%
%
%\vspace{1em}
%
%
%
%\vspace{1em}
%
%
%
%\vspace{1em}
%
%

\section*{Acknowledgment}

This research was supported in part by the Intramural Research Programs of the National Library of Medicine (NLM) and National Institutes of Health (NIH) Clinical Center. It also was supported by NLM under Award No. 4R00LM013001. This work utilized the computational resources of the NIH HPC Biowulf cluster (\url{http://hpc.nih.gov}). This material is also based upon work supported by Google Cloud. 

%\section*{References}

%%Harvard
\bibliographystyle{abbrvnat}
\bibliography{refs}

\end{document}